\documentclass[aps,pra,twocolumn,superscriptaddress,shownopacs]{revtex4-1}

\usepackage{graphicx}
\usepackage{subfigure}
\usepackage{hyperref}
\usepackage{amsmath}
\usepackage{amssymb}
\usepackage{epstopdf}
\usepackage{bbm}

\begin{document}

\title{Slowing Quantum Decoherence by Squeezing in Phase Space}

\author{H. Le Jeannic}
\affiliation{Laboratoire Kastler Brossel, Sorbonne Universit\'e, CNRS, PSL Research University, Coll\`ege de France, 4 Place
Jussieu, 75005 Paris, France}
\author{A. Cavaill\`es}
\affiliation{Laboratoire Kastler Brossel, Sorbonne Universit\'e, CNRS, PSL Research University, Coll\`ege de France, 4 Place
Jussieu, 75005 Paris, France}
\author{K. Huang}
\affiliation{Laboratoire Kastler Brossel, Sorbonne Universit\'e, CNRS, PSL Research University, Coll\`ege de France, 4 Place
Jussieu, 75005 Paris, France}
\affiliation{Shanghai Key Laboratory of Modern Optical Systems, and Engineering Research Center of Optical Instruments and Systems (Ministry of Education), School of Optical Electrical and Computer Engineering, University of Shanghai for Science and Technology, Shanghai 200093, China}
\author{R. Filip}
\email{filip@optics.upol.cz}
\affiliation{Department of Optics, Palack\'y University, 17. listopadu 1192/12, 77146 Olomouc, Czech Republic}
\author{J. Laurat}
\email{julien.laurat@sorbonne-universite.fr}
\affiliation{Laboratoire Kastler Brossel, Sorbonne Universit\'e, CNRS, PSL Research University, Coll\`ege de France, 4 Place
Jussieu, 75005 Paris, France}


\begin{abstract}
Non-Gaussian states, and specifically the paradigmatic Schr\"odinger cat state, are well-known to be very sensitive to losses. When propagating through damping channels, these states quickly loose their non-classical features and the associated negative oscillations of their Wigner function. However, by squeezing the superposition states, the decoherence process can be qualitatively changed and substantially slowed down. Here, as a first example, we experimentally observe the reduced decoherence of squeezed optical coherent-state superpositions through a lossy channel. To quantify the robustness of states, we introduce a combination of a decaying value and  a rate-of-decay of the Wigner function negativity. This work, which uses squeezing as an ancillary Gaussian resource, opens new possibilities to protect and manipulate quantum superpositions in phase space. 
\end{abstract}

 
\maketitle

The coupling of a quantum system to ambient environment leads to the so-called decoherence phenomenon \cite{ZurekRMP}. This irreversible process washes out the non-classical features of quantum states and constitutes a main impediment to quantum information sciences \cite{HarocheBook}, including for quantum computing, metrology and communication. In quantum physics, decoherence typically scales with time, strength of the coupling and also with an effective dimension of the system. Larger systems exhibit faster decoherence and are extremely fragile.

A paradigmatic example is the superposition of coherent states (CSS), also known as Schr\"odinger cat state. This state of the form $|\alpha\rangle\pm|-\alpha\rangle$ consists in a superposition of coherent states with opposite phases and mean number of energy quanta $|\alpha|^2$. In the recent years, CSS have been prepared in a variety of experimental platforms, including trapped-ion setups \cite{Monroe1996,Myatt2000, Steane2007,Home2015,Home2016}, photonics experiments \cite{Ourjoumtsev2006, Polzik2006, Sasaki2008,Ourjoumtsev2007, Etesse2015, Huang2015}, cavity-QED systems \cite{Deleglise2008}, and superconducting circuits \cite{Shoelkopf2013}. In subsequent studies, these states have been used to follow their decoherence under energy loss and to explore the boundary between the classical and quantum worlds \cite{Brune96,Myatt2000,Deleglise2008}. 

In this context, over the past several years, different strategies have been developed to mitigate the effect of decoherence and protect non-classical features. Error-correcting codes \cite{Shor1995,Steane1996}, protected logical qubits with redundant information encoding \cite{Mirrahimi2014} or probabilistic quantum distillation protocols \cite{Bennett96} are actively pursued. Physical control of decoherence via measurement and feedback has also been proposed and implemented \cite{Tombesi,Sayrin}. In all these approaches, the effect of decoherence on a given state is alleviated afterwards, without an engineering of the environment. By a challenging quantum squeezing of the environmental bath, the decoherence can be reduced as well \cite{Walls1988, Illuminati2004}. 
 
In contrast, another equivalent but potentially more feasible strategy can consist in slowing the decoherence by acting on the state itself. Such protection yet to be demonstrated can be obtained for instance via a Gaussian squeezing operation that adapts the non-Gaussian state prior to the channel \cite{Filip2001,Illuminati2004,Radim2013}. The squeezing process does not preserve energy and translates in phase space to a simple scale transformation that compresses one direction and dilates the orthogonal one. By definition, it therefore preserves the non-Gaussian content of the initial state, and in particular the negative fringes in the Wigner function representation in phase space \cite{SchleichBook}. Squeezing constitutes the essential and broadly available resource in continuous-variable quantum optics but recent proposals have also shown the usefulness of Gaussian operations on non-Gaussian states \cite{Menzies2009}. Moreover, a deterministic squeezing gate has been recently demonstrated and is now available for optical protocols \cite{Miwa2014}.

In this Letter, we report first quantitative measurements that demonstrate a decoherence much slower than the standard one typically observed for large coherent-state superpositions. Specifically, we experimentally generate a squeezed optical version of CSS and follow the behavior of the Wigner function oscillations under photon loss. We simultaneously determine the decaying value and a so-called rate-of-decay of the Wigner function negativity and compare it with standard decoherence. We demonstrate that both are substantially reduced for the squeezed CSS. Such observations therefore elevate the squeezing operation as a fundamental off-line Gaussian ressource for quantum information processing. 

\begin{figure}[t!]
\centering
\includegraphics[width=0.96\columnwidth]{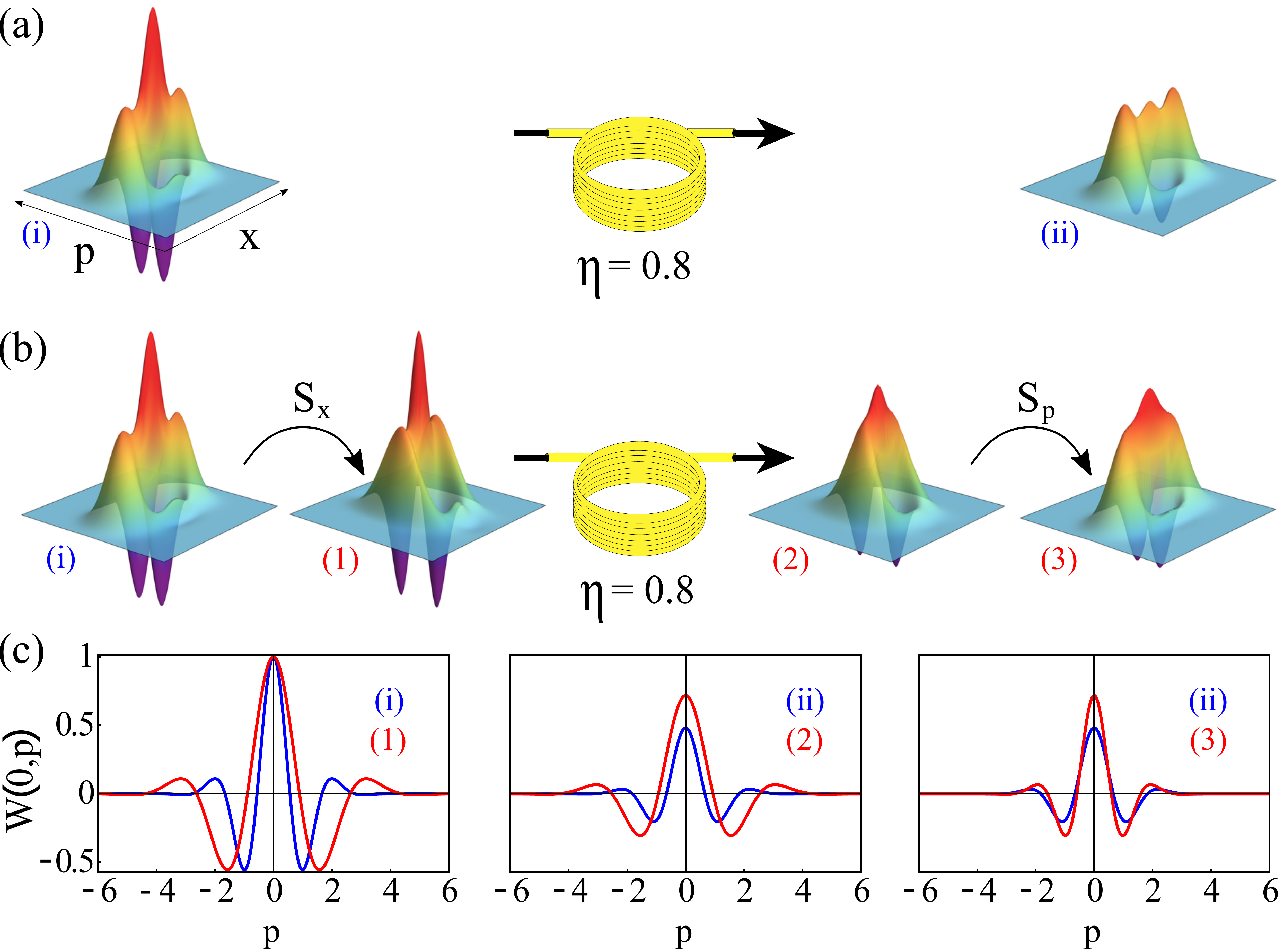}
\caption{(color online). Illustration of the decoherence process in phase space and protection by a squeezing operation. (a) Direct transmission. The theoretical Wigner functions of an even coherent-state superposition ($|\alpha|^2=2$) are displayed before and after propagation through a lossy channel with a transmission $\eta$= 0.8. (b) Squeezing adaptation before transmission. The same coherent-state superposition is first squeezed ($S_x=4$~dB) and then transmitted. An anti-squeezing operation may then be performed. (c) The 2D plots give the cross sections of the Wigner functions along the $p$ axis. The oscillations are damped with the loss but the negativities are better preserved for the squeezed version.}
\label{fig1}
\end{figure}

To first illustrate the Gaussian adaptation strategy, Fig.~\ref{fig1} depicts the specific case considered in this study. A squeezed coherent-state superposition and a CSS with same amplitude but without squeezing propagate over a lossy channel. The theoretical Wigner functions and their cross sections are given before and after the in-line attenuation. As can be seen, the Wigner function oscillations are damped with the loss but the fringe contrast and the negativities are better preserved for the squeezed superposition. If necessary, depending on the subsequent use, antisqueezing can be applied after propagation. We will not consider this additional operation in the following as it does not change the Wigner negativity values.

\textit{Gaussian adaptation.---} We begin by considering a lossy Gaussian channel with a transmission $\eta$, and a squeezing operation on the environmental bath with an amplitude gain $\sqrt{G}$ (squeezing $S$ in dB, $S=-10\log(G)$). After the channel, the non-commuting operators corresponding to the phase-space variables are transformed following the relations:
\begin{eqnarray}
X_{out}&=&\sqrt{\eta}X_{in}+\sqrt{1-\eta}\sqrt{G}X_{env}\nonumber\\
P_{out}&=&\sqrt{\eta}P_{in}+\sqrt{1-\eta}\sqrt{G^{-1}}P_{env}.
\end{eqnarray}
The uncorrelated operators $X_{env}$ and $P_{env}$ stand for the environmental noise and the correlation between them can always be cancelled by properly changing the phase before and after the channel. By combining squeezing and phase shift, this noise can be therefore arbitrarily cancelled in any direction of the phase space at the cost of an increase in the perpendicular direction.  From these expressions, it also follows that the strategy consisting in squeezing the initial state is {\em equivalent} to the challenging squeezing of the environmental noise, as suggested in Ref. \cite{Illuminati2004}. We note however that in \cite{Illuminati2004}, the decoherence is studied via the evolution of the state purity. However, purity as well as fidelity are parameters that average all features of the considered state and do not directly relate to the important nonclassical aspects of the interference in phase space. 

Turning to the phase-space representation, the Wigner function $W(x,p)$ of the initial state is thereby transformed to $W'(x',p')$ at the channel output by a two-dimensional convolution with a kernel depending on the environment \cite{SchleichBook} 
\begin{align}
&K(x,p,x',p')\propto \nonumber\\ 
&\exp\left(-\frac{(x'-\sqrt{\eta}x)^2}{2G(1-\eta)V_{x,env}} -\frac{(p'-\sqrt{\eta}p)^2}{2G^{-1}(1-\eta)V_{p,env}} \right),
\label{eq2}
\end{align}
where $V_{x,env}$ and $V_{p,env}$ are variances of the environmental noise, respectively. Due to the global nature of the convolution in the phase space, the value and direction of the squeezing has to be optimized separately for different input states $W(x,p)$ and for different measures of decoherence.

Using Eq.(\ref{eq2}) for purely lossy channel, the output Wigner function $W'(0,0)$ is clearly non-negative for any $\eta<1$ if the gain $G$ asymptotically tends to zero or diverges to infinity. Simultaneously, for the states with $W(0,0)<0$ and a non-vanishing derivative $dW'(0,0)/dG|_{G=1}$, a squeezing along the $x$ or $p$ quadratures helps to slow down decoherence. As a result, there exists an optimal squeezing value. The derivative gives a sufficient integral condition to reach it. Moreover, as any negativity of $W(x,p)$ can be shifted to the origin, this analysis applies therefore to many asymmetrical states in phase space. The simplest examples have been theoretically presented in \cite{Radim2013} and optimally-squeezed CSS are derived in the Appendix. 

\begin{figure}[b!]
\centering
\includegraphics[width=0.9\columnwidth]{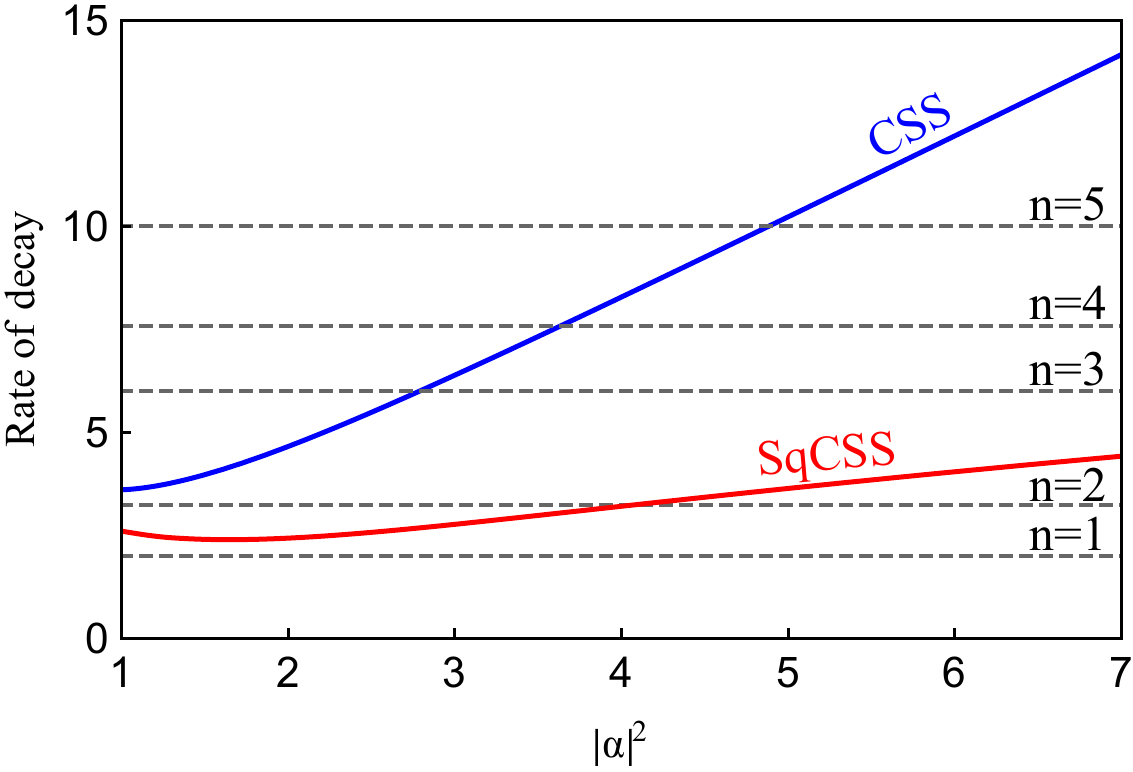}
\caption{(color online). Protection of coherent-state superpositions by squeezing. The rates of decay at $\eta=1$ for CSS (blue solid line) and optimally-squeezed CSS (red solid line) are plotted as a function of the mean photon number $|\alpha|^2$. For comparison, the horizontal grey dashed lines give the rate of decay for Fock states $|n\rangle$.}
\label{fig2}
\end{figure}

\textit{Decoherence quantification.---} We now turn to the quantification of decoherence. Any modulation of the Wigner function into the highly-nonclassical \textit{negative} values gradually decays under increasing dissipation to a zero-temperature environment \cite{Paz1993}. The presence of such negative values is a clear witness of nonclassical phase space interference. However, it is not sufficient to describe the dynamics of decoherence in the channel. 

For this purpose, we adapt the approach initiated by W.H. Zurek and coworkers using a relative rate $\dot{W}(x,p)/W(x,p)$, where the dot denotes time derivative in \cite{Paz1993}. Our steps beyond are the following. First, the derivative can be respective to any parametrization of the decay. In our case, it is the channel transmission $\eta$. Second, we focus only on the decay of {\em negative} values of the Wigner function. Third, we specifically measure the rate of the decay of the largest negativity located at the coordinate $x_{min}$ and momentum $p_{min}$ at the given transmission $\eta$. Therefore, we use the rate-of-decay defined as: 
\begin{equation}
RD_{W_\eta}=\frac{1}{W(x_{min},p_{min},\eta)} \frac{\partial W(x,p,\eta')}{\partial \eta' }\biggr\rvert_{x_{min},p_{min},\eta}.
\end{equation}

In Fig. \ref{fig2}, we present the evolution of the rate-of-decay for a CSS with a size $|\alpha|^2$, at a transmission $\eta=1$, as well as the value for several Fock states for comparison. For the Fock states $|1\rangle$ and $|2\rangle$, the rate-of-decay is equal to $RD_{|1\rangle}~=~2$ and $RD_{|2\rangle}=3.22$ for $\eta=1$, demonstrating that the decay is faster for the higher Fock states. More generally, for any two Fock states $|n\rangle$ and $|n+1\rangle$, we obtain $RD_{|n+1\rangle}>RD_{|n\rangle}$. For CSSs with different amplitudes, we have as well $RD(\alpha+\Delta \alpha)>RD(\alpha)$. For large mean photon number, Fock states and CCSs tend to have a comparable rate-of-decay, which increases with the size of the system (see Appendix). As illustrated in Fig. \ref{fig2}, squeezing a CSS enables to strongly reduce the rate of decay and to make it comparable to the one of smaller Fock states. This reduction is even greater for CSS with large size \cite{Radim2013}.

\textit{Experimental generation of squeezed optical CSS.---} We proceed to the experimental investigation. We generate free-propagating squeezed even CSSs via our recently-demonstrated versatile method based on two-photon heralding \cite{Huang2015}. This scheme relies on a two-mode squeezed vacuum state emitted by a type-II optical parametric oscillator (OPO) operated far below threshold \cite{Morin2012} and linear mixing of the two entangled modes (see Appendix). A two-photon detection based on high-efficiency superconducting nanowire single-photon detectors \cite{LeJeannic2016} on one of the resulting modes heralds the generation of a superposition of zero and two-photon Fock states on the other. This superposition can be tuned with the adjustable linear mixing and shows large fidelity with the targeted squeezed CSS for $|\alpha|^2$ as large as 3. Experimental details have been reported elsewhere \cite{Huang2015,Huang2016}.  

\begin{figure}[t!]
\centering
\includegraphics[width=0.95\columnwidth]{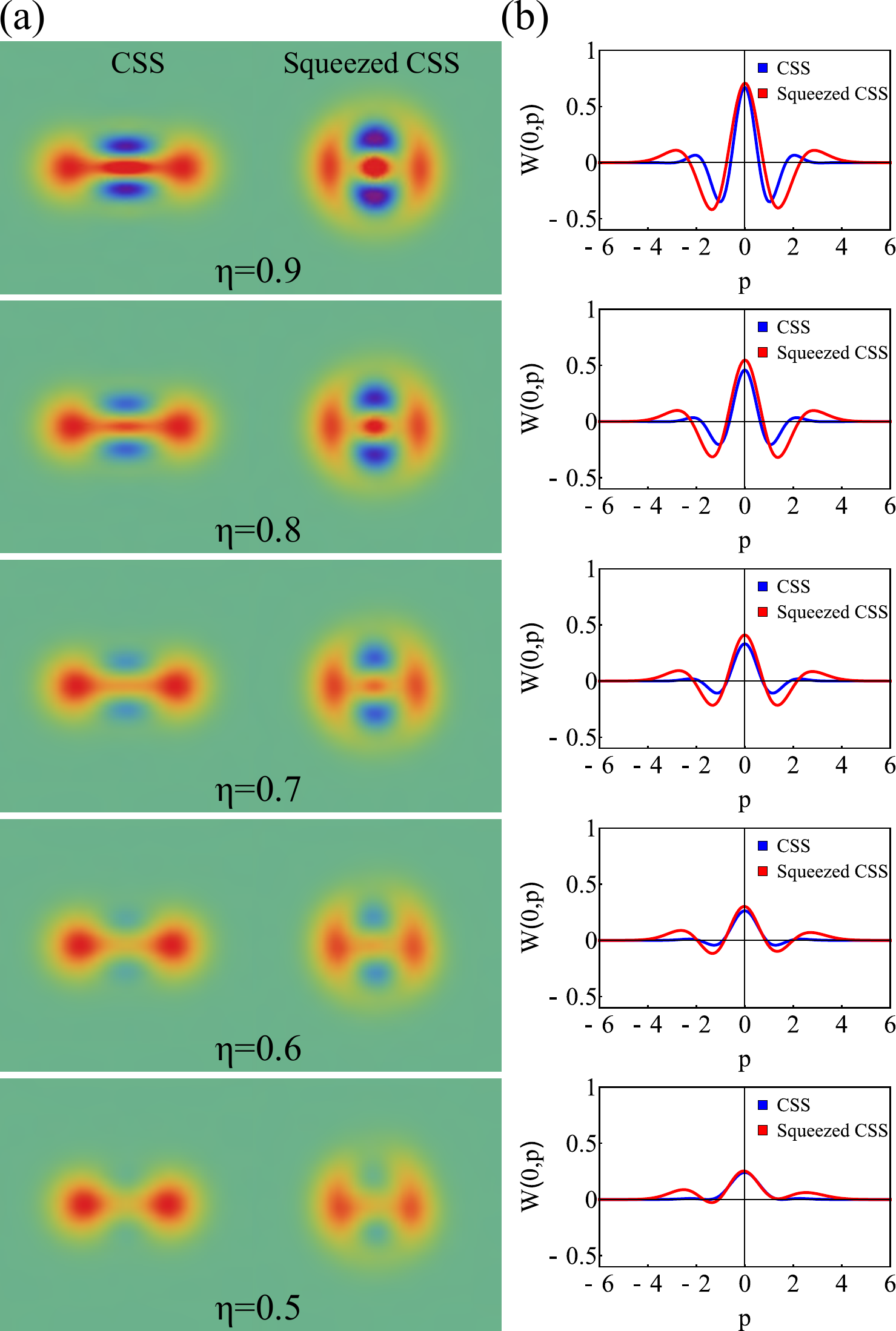}
\caption{(color online). Experimental results. (a) The Wigner functions of the experimental squeezed optical CSS are given for five different values of channel transmission $\eta$ and compared with a CSS of same amplitude without squeezing. The initial synthesized state exhibits a fidelity $F=77\%$ with a squeezed CSS with $|\alpha|^2=2.1$ and a squeezing value $S=4$~dB. (b) The right column provides the cross sections along the imaginary axes. The initial state corresponding to $\eta=0.9$ is corrected for 15\% detection losses. The specific value $\eta=0.9$ corresponds to the estimated escape efficiency of the OPO.}
\label{fig3}
\end{figure}

The heralded state is characterized by full quantum state tomography via homodyne detection. Quadrature values from 50 000 measurements are processed with a maximum likelihood algorithm \cite{Raymer}, which provides the density matrix and the associated Wigner function. The Wigner function is given in Fig. \ref{fig3} (panel $\eta=0.9$), after correction for detection losses (15\%). The state exhibits a fidelity $F\sim77\%$ with a squeezed CSS with $|\alpha|^2~=~2.1$ and $S~=~4$ dB. The negative peaks of the Wigner functions reach $W=-0.32\pm0.01$. 

The aim of the present investigation is to study the effect of losses on this squeezed CSS. For this purpose, we introduce losses by changing the temporal mode used for the homodyne characterization \cite{Morin2013,Morin2013b}. Indeed, because of the continuous-wave nature of our experiment, the state reconstruction requires a temporal filtering. The temporal mode is given by a double-decaying exponential profile $f(t)=\sqrt{\pi\gamma}e^{-\pi\gamma|t|}$, where $\gamma$ is the bandwidth of the OPO cavity. By mismatching the temporal mode and the optimal one, i.e. by adding a delay $\tau$ in the reconstruction algorithm, we introduce controllable loss, with an effective overall transmission $\eta$. This procedure works because the orthogonal modes are very close to the vacuum state as the OPO is pumped far below threshold. It provides the states after the lossy channel and these decohered states are used in the following of the study.

\textit{Experimental decoherence for squeezed optical CSS.---}
The evolution of the Wigner function under decoherence is compared for the experimentally prepared squeezed CSS and for the CSS of same amplitude but without squeezing. Figure \ref{fig3}(a) shows the Wigner functions for five different values of channel transmission $\eta$ and Fig. \ref{fig3}(b) provides the corresponding cross sections along the imaginary axis. Decoherence manifests in the progressive reduction of the Wigner function negativities. This comparison confirms that these negativities are better preserved for the squeezed CSS. 

\begin{figure}[t!]
\centering
\includegraphics[width=0.79\columnwidth]{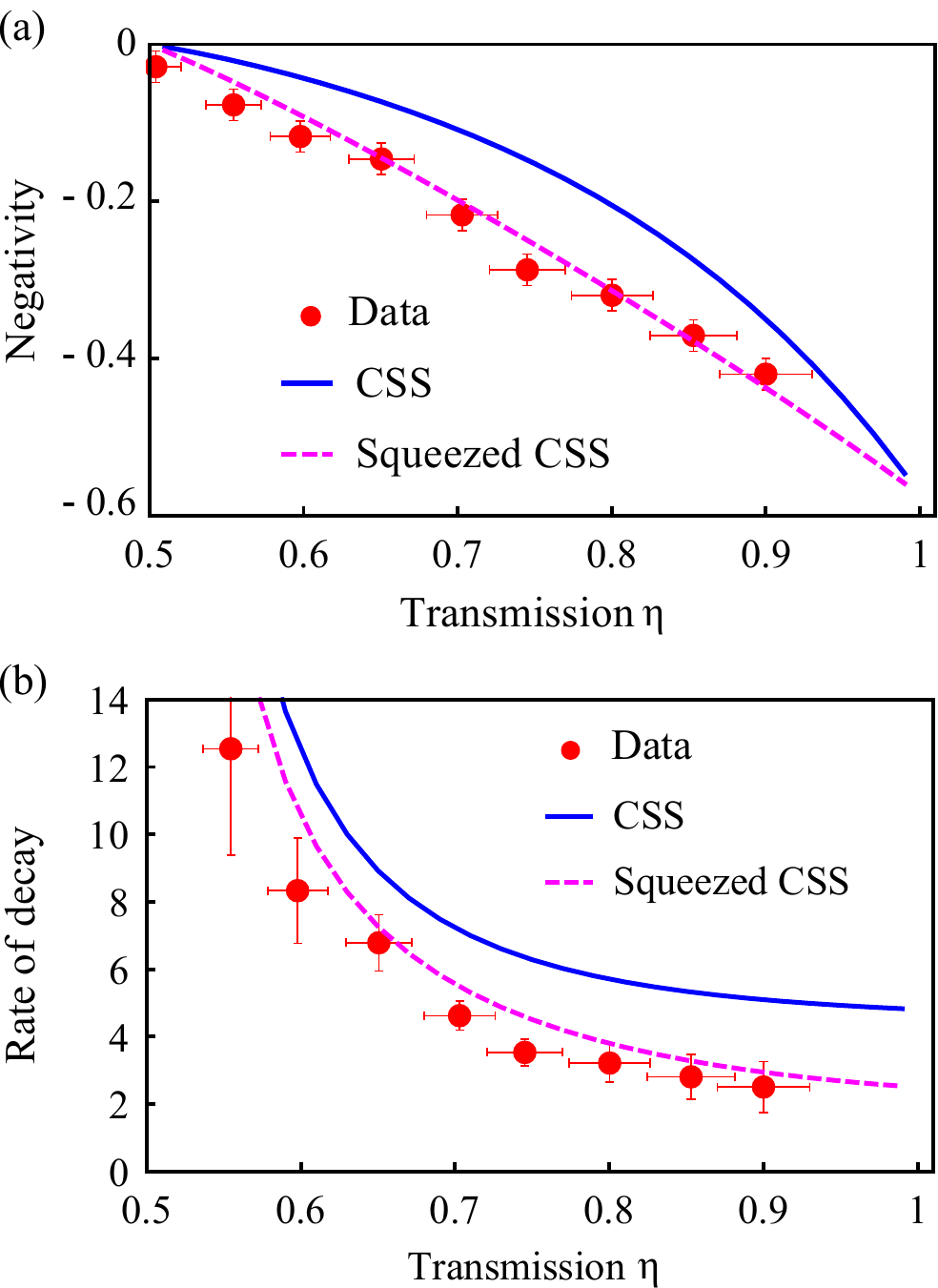}
\caption{(color online). Evolution of the Wigner function negativity and rate-of-decay as a function of the transmission $\eta$. (a) Negativity of the Wigner function of the initial squeezed CSS with $|\alpha|^2=2.1$ and $S=4$ dB. (b) Estimated rate of decay of the negativity. The solid lines show the evolution for the CSS of same amplitude without squeezing. The dashed lines correspond to the theoretical model.}
\label{fig4}
\end{figure}

More quantitatively, Fig. \ref{fig4}(a) provides the maximal Wigner function negativity as a function of the transmission $\eta$. We see that the negativity stays larger for the squeezed superposition, whatever the amount of loss experienced by the state. By fitting the experimental points by a third-order polynomial, we can finally estimate the derivative of the negativity, and therefore the rate-of-decay as a function of the transmission $\eta$, as shown in Fig. \ref{fig4}(b). The protection simultaneously improves both the negativity and the rate-of-decay, for any losses. For example, for losses around 20\%, the rate-of-decay is typically decreased by about a factor 2, a very substantial improvement. The pink dashed lines correspond to a simple theoretical model taking into account the mean-photon number and the squeezing value. The small discrepancy is explained by the higher photon-number components included in this model and not present in our experimentally synthesized superposition.

\textit{Discussion.---} 
Our result clearly illustrates the decoherence protection provided by the squeezing adaptation of fragile non-classical feature in the coherent-state superpositions. This improved robustness can find immediate applications in quantum communication and computing schemes. It has been shown for instance that CSS code-words can be used for quantum memory protection \cite{Leghtas2013} and this study has been extended to multicomponent superpositions \cite{Loock2016}. The use of squeezed code-words may reduce further the error rate. In a similar way, recent demonstrations based on CSS for universal gate implementations in superconducting circuits \cite{Heeres2017} could benefit from this approach to further limit errors. Protecting non-classical features in experiments where such states are used to study various topological effects is also highly relevant \cite{Flurin2017}. The general idea to optimize CSS can be analyzed for a variety of other cat-like states used in different applications proposed sometimes long-time ago, such as entangled CSS for quantum repeaters \cite{Loock2006} or compass states for quantum metrology \cite{Toscano2006}. 

Beyond cat-like states, this strategy can be applied to protect other quantum states. For lossy channels, it is especially powerful for the phase-asymmetrical states that exhibit complex topology of negativity in phase space \cite{Furusawacomplex,Hofheinz2009}. These local operations may also provide advantages for protecting complex bipartite states, including NOON states or hybrid entangled states \cite{Morin2014,Jeong2014}. However, as stated before, an exact prediction for a specific state must be analyzed always relatively to a given figure of merit. The channel is described by a convolution of the input Wigner function with a kernel representing the channel noise. As a nonlocal transformation in phase space, the convolution forbids for a general state any simple intuitive picture of how and why the squeezing helps. It is not exactly related to minimization of energy or to azimuthal symmetry in phase space. Importantly, the nonlocal nature of this transformation means involvement of all small details of the Wigner function in the decoherence process. A general operational procedure is therefore dependent not only on the specific state but also on the feature to protect for a given application.
 
The squeezing adaptation used here can be further generalized to any Gaussian operations. If we accept a probabilistic protocol, any conditional purity-preserving Gaussian filter can be added to the toolbox to protect against decoherence \cite{Jaromir}. For multimode states, useful for example in quantum coding, multimode Gaussian operations and filters can also be advantageous. Beyond the lossy channel, the method is extendable to noisy and non-Gaussian channels \cite{Hage2008}.

In conclusion, we have shown that the photon-loss induced decoherence of CSS can be significantly slowed down by initially squeezing the superposed states. This strategy allows here to reduce the rate of decay by a factor 2. Theory remarkably predicts even higher factors for larger superpositions. Our study thereby confirms the usefulness of Gaussian operations on non-Gaussian states to make them more robust and to approach a minimal decoherence rate. This Gaussian adaptation is a versatile tool for advanced quantum state engineering and robust state transfer, and can find immediate applications in a variety of schemes. This work also invites to revisit the standard decoherence rate usually considered for many other physical platforms, where quantum superpositions of macroscopic states can be generated. In a broader context, this result constitutes a further example of the potential of the hybrid quantum information processing approach where Gaussian and non-Gaussian operations and techniques can be combined for advanced capabilities and improved scalability \cite{Andersen}.

\begin{acknowledgments}
The authors thank O. Morin, Y.-C. Jeong and J. Ruaudel for their contributions in the early stage of the experiment. This work was supported by the European Research Council (Starting Grant HybridNet). R.F. also acknowledges funding by the grant GB14-36681G of the Czech Science Foundation. J.L. is a member of the Institut Universitaire de France.
\end{acknowledgments}

\clearpage

\onecolumngrid
\appendix

\section{Experimental setup}
\begin{figure}[b!]
\centering
\includegraphics[width=0.55\columnwidth]{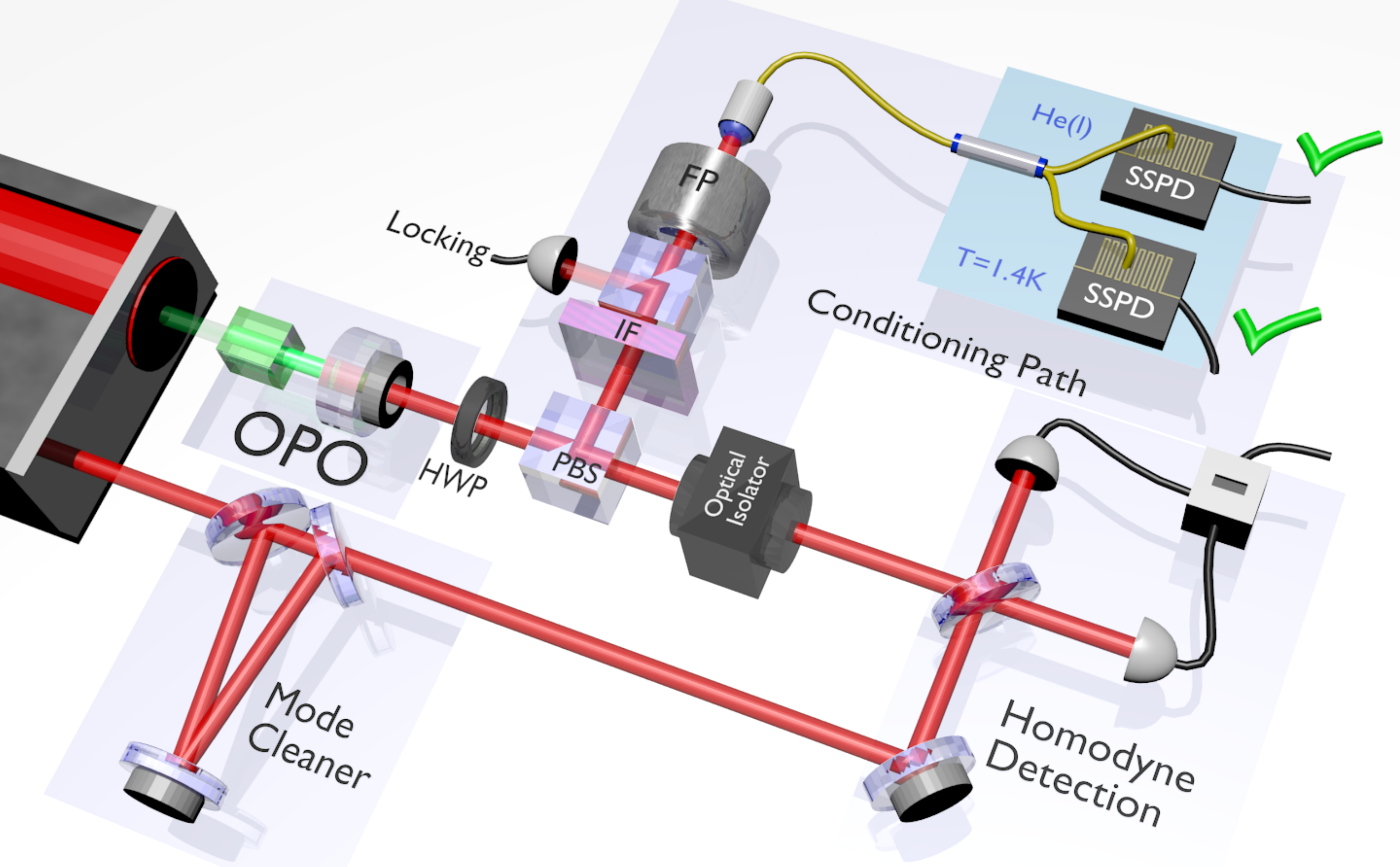}
\caption{Experimental setup for the generation of heralded squeezed coherent-state superpositions. It is based on a two-mode squeezed vacuum generated by a type-II OPO operated below threshold. The two orthogonally-polarized signal and idler modes are slightly mixed via a half wave plate and a polarizing beam splitter. A two-photon detection on the conditioning path heralds the generation of a squeezed CSS, which is then characterized via homodyne detection.}
\label{figure1}
\end{figure}

The experimental setup is sketched in Fig. \ref{figure1}. It consists of a continuous-wave type-II phase-matched optical parametric oscillator made of a semi-monolithic linear cavity and pumped far below threshold. On one side, a 1cm-long KTP crystal is high-reflection-coated at 1064 nm and $R=95\%$ at the 532 nm pump. The output coupler is made of a curved mirror HR-coated for the pump and with a reflection $R=90\%$ at 1064 nm \cite{Morin2012}.

The orthogonally-polarized signal and idler modes are mixed using a polarizing beam-splitter (PBS) and a half-wave plate (HWP). A two-photon detection is implemented on the conditioning path, on one of the outputs of the polarizing beam-splitter after frequency filtering, by the use of two superconducting nanowire single-photon detectors \cite{LeJeannic2016}. The detection of coincidence events in an acceptance window smaller than 1~ns heralds the generation. This acceptance window is much smaller than the temporal mode duration ($\sim 20$ ns) of the emitted photons, which is given by the inverse of the OPO bandwidth \cite{Morin2013}. The heralded state is then characterized by quantum state tomography via homodyne detection.

If the signal and idler modes are perfectly separated, two-photon detection events herald the generation of two-photon Fock states. However, if one adds a small angle to the half-wave plate before the polarizing beam-splitter, states exhibiting a high fidelity with squeezed even coherent-state superpositions are heralded \cite{Huang2015}. 
 
\section{Wigner function negativity and rate of decay}
In the following, we give the expressions for the Wigner function negativity and the corresponding rate of decay as a function of the channel transmission $\eta$ for different quantum states. 

\begin{itemize}
\item Fock states $|n\rangle$. The rate of decay depends on the photon number.
For a single-photon state, the negativity of the Wigner function corresponds to its value at the origin of the phase space. This negativity decreases linearly with the channel loss. 
\begin{eqnarray}
W_{|1\rangle}(0,0)&=&1-2 \eta \nonumber\\
RD_{|1\rangle}(\eta)&=&\frac{2}{2 \eta-1}\nonumber
\end{eqnarray}
For a transmission $\eta=1$, the rate of decay is $RD_{|1\rangle}=2$.\\

When the photon number increases, the rate of decay also increases.
For example, for $n=2$, the analytical expressions can be written as:
\begin{eqnarray}
W_{|2\rangle}(0,p_{\textrm{min}})&=&-2e^{\frac{\sqrt{12 \eta ^2-8 \eta +2}-6 \eta +2}{2 \eta }} \eta  \left(\sqrt{12 \eta ^2-8
   \eta +2}-2 \eta \right)\nonumber\\
   RD_{W_{|2\rangle}}(\eta)&=&\frac{(2 \eta -1) \left(\sqrt{12 \eta ^2-8 \eta +2}-2 \eta +1\right)}{\eta ^2 \left(\sqrt{12 \eta
   ^2-8 \eta +2}-2 \eta \right)}\nonumber
\end{eqnarray}
with $p_{\textrm{min}}=\sqrt{-\frac{\sqrt{12 \eta ^2-8 \eta +2}-6 \eta +2}{\eta }}$.\\
For a transmission $\eta=1$, the rate of decay is $RD_{|2\rangle}=\frac{1}{2}(4+\sqrt{6})\approx 3.22$.

The rates of decay of several Fock states are shown in Fig. \ref{figure2}(a) as a function of the channel transmission $\eta$. It increases monotonically with $n$. 

\begin{figure}[b!]
\centering
\includegraphics[width=0.6\columnwidth]{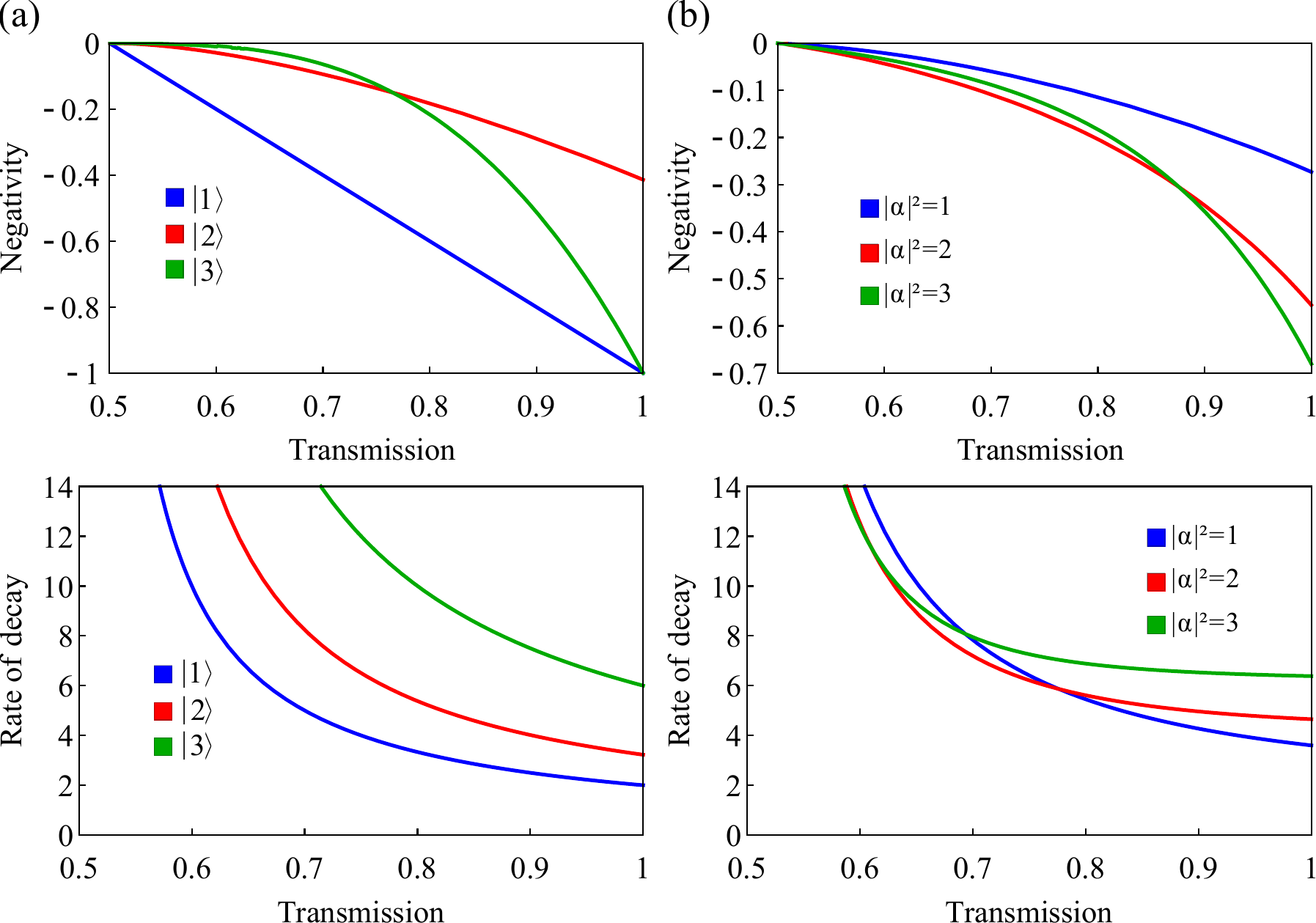}
\caption{Wigner function negativity and rate of decay for (a) Fock states and for (b) coherent-state superpositions with different mean photon numbers $|\alpha|^2$, as a function of the channel transmission $\eta$. }
\label{figure2}
\end{figure}
\item Coherent-state superposition $|\textrm{CSS}\rangle\propto|\alpha\rangle+|-\alpha\rangle$. 
The expression of the Wigner function for any efficiency $\eta$ can be written as:
\begin{eqnarray}
W_{|\textrm{CSS}\rangle}(x,p,\eta)=\frac{e^{-\frac{x^2+p^2}{2}-2\alpha^2(\eta-1)-2x\alpha \sqrt{\eta}}}{2  \left(e^{2 \alpha ^2}+1\right)}(2 e^{2 \alpha  \left(\alpha  (2 \eta -1)+\sqrt{\eta } x\right)} \cos \left(2
   \alpha  \sqrt{\eta } p\right)+e^{4 \alpha  \sqrt{\eta } x}+1).\nonumber
\end{eqnarray}

The rate of decay is given by:
\begin{eqnarray}
RD_{W_{(|\textrm{CSS}\rangle)}}(\eta)=-\frac{\alpha  \left(4 \alpha  \sqrt{\eta }-4 \alpha  \sqrt{\eta } e^{2 \alpha ^2 (2 \eta -1)} \cos \left(2 \alpha  \sqrt{\eta } p_\textrm{min}\right)+2 p_\textrm{min} e^{2 \alpha  (2 \alpha  \eta -\alpha )}
   \sin \left(2 \alpha  \sqrt{\eta } p_\textrm{min}\right)\right)}{\sqrt{\eta } \left(2 e^{2 \alpha ^2 (2 \eta -1)} \cos \left(2 \alpha  \sqrt{\eta } p_\textrm{min}\right)+2\right)}.\nonumber
\end{eqnarray}
For an even CSS, the rate of decay at a transmission $\eta=1$ also increases with the size $|\alpha|^2$, as shown in Fig.\ref{figure2}(b).
\end{itemize}

Note that the important regime for both the negativity of the Wigner function and the associated rate-of-decay is for $\eta$ close to unity. 
In this region lie the highly non-classical states necessary for quantum processing applications. Conversely, the behavior for low $\eta$ is not as interesting when it comes to applications to subsequent protocols.  
\begin{figure}[t!]
\centering
\includegraphics[width=0.4\columnwidth]{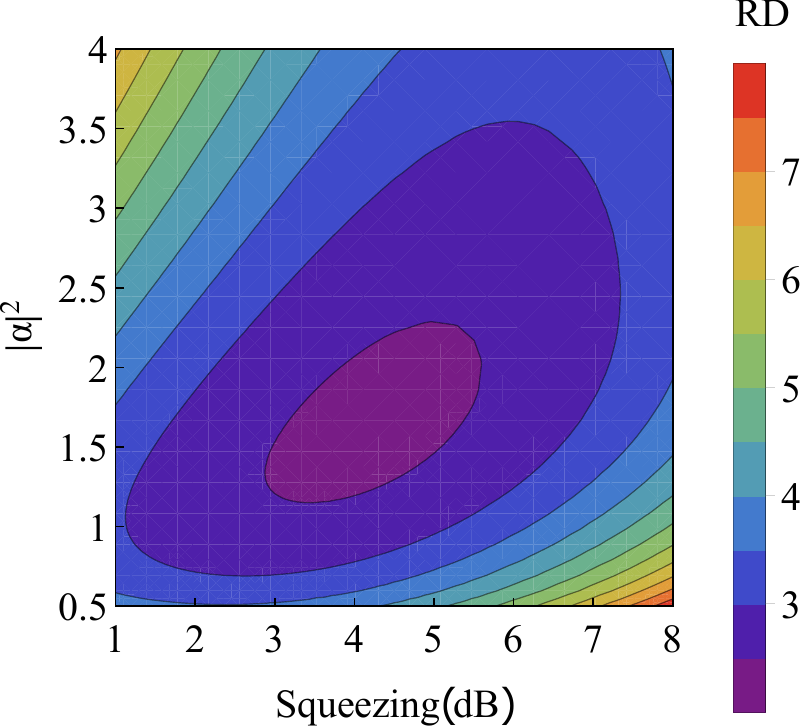}
\caption{Rate of decay for squeezed CSSs as a function of the mean photon number $|\alpha|^2$ and the squeezing value, at a perfect transmission $\eta=1$. }
\label{figure3}
\end{figure}

\section{Optimally squeezed coherent-state superposition}
In order to reduce the rate of decay, it is possible to squeeze the coherent-state superposition. The Wigner function of a squeezed CSS can be written as:
\begin{eqnarray}
W_{|CSS\rangle}(x,p) \rightarrow W_{\hat{S}(\xi)|\textrm{CSS}\rangle}(x,p,\xi)=W_{|\textrm{CSS}\rangle}(x e^{\xi},p e^{-\xi})\nonumber
\end{eqnarray}
By denoting $G=e^{-2\xi}$ where $\xi$ is the squeezing parameter, one obtain:
\begin{multline}
W_{\hat{S}|\textrm{CSS}\rangle}(x,p,G)=\frac{e^{-\frac{p^2 G^2+4 \alpha  \sqrt{G} x+x^2}{2 G}}}{2  \left(e^{2 \alpha ^2}+1\right)} \times \\
 \left(2 \cos \left(2 \alpha  p \sqrt{G}\right) \sinh \left(2 \alpha  \left(\alpha +\frac{x}{\sqrt{G}}\right)\right)+2 \cos \left(2 \alpha  p \sqrt{G}\right) \cosh
   \left(2 \alpha  \left(\alpha +\frac{x}{\sqrt{G}}\right)\right)+\sinh \left(\frac{4 \alpha  x}{\sqrt{G}}\right)+\cosh \left(\frac{4 \alpha  x}{\sqrt{G}}\right)+1\right)\nonumber
\end{multline}
Then, after a given transmission $\eta$:
\begin{multline}
W_{\hat{S}|\textrm{CSS}\rangle}(x,p,G)\times W_{|0\rangle_\textrm{E}}(x_\textrm{E},p_\textrm{E}) \rightarrow  W_{\textrm{Tr}[\hat{B}(\eta)\hat{S}(G)|\textrm{CSS}\rangle\otimes |0\rangle_\textrm{E} \langle 0 | \langle \textrm{CSS} | ...]_\textrm{E}}(x,p,G,\eta)\\
=\iint W_{\hat{S}|\textrm{CSS}\rangle}(\sqrt{\eta} x +\sqrt{1-\eta} x_\textrm{E},\sqrt{\eta} p +\sqrt{1-\eta} p_\textrm{E},G)\times W_{|0\rangle_\textrm{E}}(\sqrt{\eta} x_\textrm{E} +\sqrt{1-\eta} x,\sqrt{\eta} p_\textrm{E} +\sqrt{1-\eta} p) dx_\textrm{E} dp_\textrm{E}\nonumber
\end{multline}
where $\hat{B}(\eta)$ is a beam splitter operation corresponding to a transmission coefficient $t=\sqrt{\eta}$. E is the environment mode, and $W_{|0\rangle_\textrm{E}}(x_\textrm{E},p_\textrm{E})=e^{-\frac{1}{2}(x_\textrm{E}^2+p_\textrm{E}^2)}$ is the Wigner function of the vacuum mode E.\\

This operation leads to the final formula for the Wigner function of a squeezed CSS. It can be written as: 
\begin{eqnarray}
W_{\hat{S}|\textrm{CSS}\rangle}(x,p,G,\eta)=\frac{A\times \left(e^{\frac{2 \alpha  G \left(\eta ^{3/2} G^2 x+\alpha  (2 \eta -1) G-(\eta -1) \sqrt{\eta } x\right)}{\left((\eta -1) G^2-\eta \right) \left(\eta  \left(G^2-1\right)+1\right)}}(e^{\frac{4 \alpha  \sqrt{\eta } G x}{\eta -\eta  G^2+G^2}}+1)+2 \cos \left(\frac{2 \alpha  \sqrt{\eta } p G}{\eta  \left(G^2-1\right)+1}\right)\right)}{2  \left(e^{2 \alpha ^2}+1\right) \sqrt{\eta +\frac{1-\eta }{G^2}} \sqrt{\eta -(\eta -1)
   G^2}}\nonumber
\end{eqnarray}
where $A=e^{\frac{p^2 \left(\eta -(\eta -1) G^2\right)+G^2 \left(4 \alpha ^2 \eta  \left((\eta -1) G^2-\eta \right)+x^2 \left(\eta  \left(G^2-1\right)+1\right)\right)}{2 \left((\eta -1) G^2-\eta
   \right) \left(\eta  \left(G^2-1\right)+1\right)}}$.\\

The expression for the rate of decay can then be obtained from the above equation. Depending on the applied squeezing amount and mean photon number of the CSS, the rate of decay  for $\eta=1$ scales as shown in Fig. \ref{figure3}. By optimizing the squeezing, it is possible to target a minimal rate of decay. The optimally-minimized rate of decay is not necessarily the same for each mean photon number of the CSS, and does not depend linearly on $|\alpha|^2$. However, an optimal squeezing that reduces the decoherence rate can always be found.

\begin{figure}[h!]
\centering
\includegraphics[width=0.8\columnwidth]{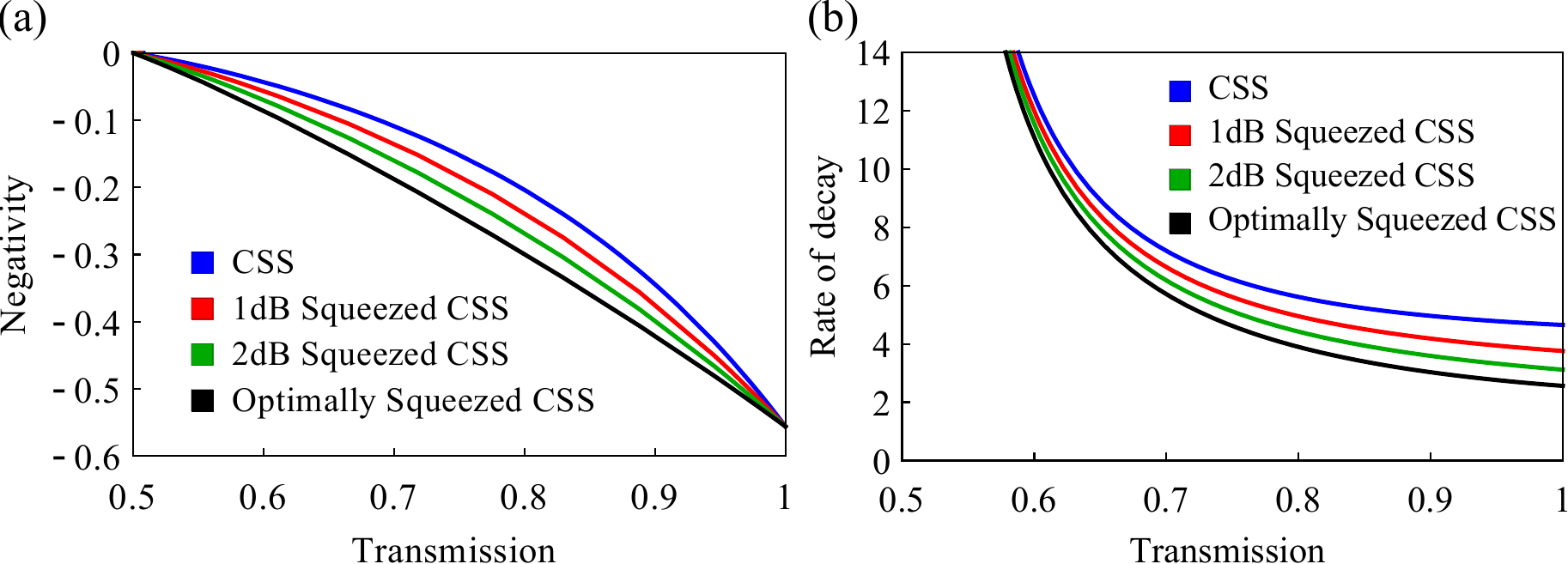}
\caption{(a) Negativity of the Wigner function and (b) rate of decay for CSS and squeezed CSS with mean photon number $|\alpha|^2=2$, as a function of the transmission $\eta$ of the channel. The evolutions are given for different squeezing values. The optimally-squeezed CSS is shown in black solid line.}
\label{figure4}
\end{figure}

For a given mean photon number, the optimal squeezing remains the same, independently of the channel efficiency, as shown in Fig. \ref{figure4} that gives the negativity of the Wigner function and the rate of decay for different squeezing values.

\end{thebibliography}


\begin{thebibliography}{99}

\bibitem{ZurekRMP} W. H. Zurek, Decoherence, einselection, and the quantum origins of the classical, Rev. Mod. Phys. \textbf{75}, 715 (2003).
\bibitem{HarocheBook} S. Haroche and J.-M. Raimond, \textit{Exploring the quantum: atoms, cavities and photons} (Oxford University Press, Oxford, 2006).

\bibitem{Monroe1996} C. Monroe, D. M. Meekhof, B. E. King, and D. J. Wineland, A Schr\"odinger Cat Superposition State of an Atom, Science \textbf{272}, 1131 (1996).
\bibitem{Myatt2000} C. J. Myatt, B. E. King, Q. A. Turchette, C. A. Sackett, D. Kielpinski, W. M. Itano, C. Monroe, and D. J. Wineland, Decoherence of quantum superpositions through coupling to engineered reservoirs, Nature \textbf{403}, 269 (2000).
\bibitem{Steane2007} M. J. McDonnell \textit{et al.}, Long-Lived Mesoscopic Entanglement outside the Lamb-Dicke Regime, Phys. Rev. Lett. \textbf{98}, 063603 (2007).
\bibitem{Home2015} H.-Y. Lo, D. Kienzler, L. de Clercq, M. Marinelli, V. Negnevitsky, B. C. Keitch, and J. P. Home, Spin motion entanglement and state diagnosis with squeezed oscillator wavepackets, Nature \textbf{521}, 336 (2015).
\bibitem{Home2016} D. Kienzler, C. Fl\"uhmann, V. Negnevitsky, H.-Y. Lo, M. Marinelli, D. Nadlinger, and J. P. Home, Observation of Quantum Interference between Separated Mechanical Oscillator Wave Packets, Phys. Rev. Lett. \textbf{116}, 140402 (2016).

\bibitem{Ourjoumtsev2006} A. Ourjoumtsev, R. Tualle-Brouri, J. Laurat, and Ph. Grangier, Generating optical Schr\"odinger kittens for quantum information processing, Science \textbf{312}, 83 (2006).
\bibitem{Polzik2006} J. S. Neergaard-Nielsen, B. Melholt Nielsen, C. Hettich, K. M\o lmer, and E. S. Polzik, Generation of a Superposition of Odd Photon Number States for Quantum Information Networks, Phys. Rev. Lett. \textbf{97}, 083604 (2006).
\bibitem{Sasaki2008} H. Takahashi, K. Wakui, S. Suzuki, M. Takeoka, K. Hayasaka, A. Furusawa, and M. Sasaki, Generation of Large-Amplitude Coherent-State Superposition via Ancilla-Assisted Photon Subtraction, Phys. Rev. Lett. \textbf{101}, 233605 (2008).
\bibitem{Ourjoumtsev2007} A. Ourjoumtsev, H. Jeong, R. Tualle-Brouri, and Ph. Grangier, Generation of optical Schr\"odinger cats from photon number states, Nature \textbf{448}, 784 (2007).
\bibitem{Etesse2015} J. Etesse, M. Bouillard, B. Kanseri, and R. Tualle-Brouri, Experimental Generation of Squeezed Cat States with an Operation Allowing Iterative Growth, Phys. Rev. Lett. \textbf{114}, 193602 (2015).
\bibitem{Huang2015} K. Huang \textit{et al.}, Optical Synthesis of Large-Amplitude Squeezed Coherent-State Superpositions with Minimal Resources, Phys. Rev. Lett. \textbf{115}, 023602 (2015).

\bibitem{Deleglise2008} S. Del\'eglise \textit{et al.}, Reconstruction of non-classical cavity field states with snapshots of their decoherence, Nature \textbf{455}, 510 (2008).

\bibitem{Shoelkopf2013} B. Vlastakis, G. Kirchmair, Z. Leghtas, S. E. Nigg, L. Frunzio, S. M. Girvin, M. Mirrahimi, M. H. Devoret, and R. J. Schoelkopf, Deterministically Encoding Quantum Information Using 100-Photon Schr\"odinger Cat States, Science \textbf{342}, 607 (2013). 
\bibitem{Brune96} M. Brune, E. Hagley, J. Dreyer, X. Ma\^itre, A. Maali, C. Wunderlich, J. M. Raimond, and S. Haroche, Observing the Progressive Decoherence of the ``Meter'' in a Quantum Measurement, Phys. Rev. Lett. \textbf{77}, 4887 (1996).

\bibitem{Shor1995} P. W. Shor, Scheme for reducing decoherence in quantum computer memory, Phys. Rev. A \textbf{52}, R2493 (1995).
\bibitem{Steane1996} A. Steane, Error correcting codes in quantum theory, Phys. Rev. Lett. \textbf{77}, 793 (1996).
\bibitem{Mirrahimi2014} M. Mirrahimi \textit{et al.}, Dynamically protected cat-qubits: a new paradigm for universal quantum computation, New J. Phys. \textbf{16}, 045014 (2014).
\bibitem{Bennett96} C. H. Bennett, D. P. DiVincenzo, J. A. Smolin, and W. K. Wootters, Mixed-state entanglement and quantum error correction, Phys. Rev. A \textbf{54}, 3824 (1996).

\bibitem{Tombesi} D. Vitali, P. Tombesi, and G. J. Milburn, Controlling the Decoherence of a ``Meter'' via Stroboscopic Feedback, Phys. Rev. Lett. \textbf{79}, 2442 (1997).
\bibitem{Sayrin} C. Sayrin \textit{et al.}, Real-time quantum feedback prepares and stabilizes photon number states, Nature \textbf{477}, 73 (2011).

\bibitem{Walls1988} T. A. B. Kennedy and D. F. Walls, Squeezed quantum fluctuations and macroscopic quantum coherence, Phys. Rev. A \textbf{37}, 152 (1988).
\bibitem{Illuminati2004} A. Serafini, S. De Siena, F. Illuminati, and M. G. A. Paris, Minimum decoherence cat-like states in Gaussian noisy channels, J. Opt. B \textbf{6}, S591 (2004).  

\bibitem{Filip2001} R. Filip, Amplification of Schr\"odinger-cat state in a degenerate optical parametric amplifier,  J. Opt. B: Quantum Semiclass. Opt. \textbf{3} S1 (2001). 
\bibitem{Radim2013} R. Filip, Gaussian quantum adaptation of non-Gaussian states for a lossy channel, Phys. Rev. A \textbf{87}, 042308 (2013).

\bibitem{SchleichBook} W. P. Schleich, \textit{Quantum Optics in Phase Space} (Wiley, Berlin, 2001).

\bibitem{Menzies2009} D. Menzies and R. Filip, Gaussian-optimized preparation of non-Gaussian pure states, Phys. Rev. A \textbf{79}, 012313 (2009).
\bibitem{Miwa2014} Y. Miwa \textit{et al.}, Exploring a new regime for processing optical qubits: squeezing and unsqueezing single photons, Phys. Rev. Lett. \textbf{113}, 013601 (2014).

\bibitem{Paz1993} J. P. Paz, S. Habib, and W. H. Zurek, Reduction of the wave packet: Preferred observable and decoherence time scale, Phys. Rev. D \textbf{47}, 488 (1993).

\bibitem{Morin2012} O. Morin, V. D'Auria, C. Fabre, and J. Laurat, High-fidelity single-photon source based on a type II optical parametric oscillator, Opt. Lett. \textbf{37}, 3738 (2012).
\bibitem{LeJeannic2016} H. Le Jeannic \textit{et al.}, High-efficiency superconducting nanowire single-photon detector for quantum state engineering in the near infrared, Opt. Lett. \textbf{41}, 5341 (2016).

\bibitem{Huang2016} K. Huang \textit{et al.}, Experimental quantum state engineering with time-separated heraldings from a continuous-wave light source: a temporal-mode analysis, Phys. Rev. A \textbf{93}, 013838 (2016).
\bibitem{Raymer} A. I. Lvovsky and M. G. Raymer, Continuous-variable optical quantum-state tomography, Rev. Mod. Phys. \textbf{81}, 299 (2009).
\bibitem{Morin2013} O. Morin, C. Fabre, and J. Laurat, Experimentally accessing the temporal mode of traveling quantum light states, Phys. Rev. Lett. \textbf{111}, 213602 (2013).
\bibitem{Morin2013b} M. Ho, O. Morin, J.-D. Bancal, N. Gisin, N. Sangouard, and J. Laurat, Witnessing single-photon entanglement with local homodyne measurements: analytical bounds and robustness to losses, New J. Phys. \textbf{16}, 103035 (2013). 


\bibitem{Leghtas2013} Z. Leghtas, G. Kirchmair, B. Vlastakis, R. J. Schoelkopf, M. H. Devoret, and M. Mirrahimi, Hardware-Efficient Autonomous Quantum Memory Protection, Phys. Rev. Lett. \textbf{111}, 120501 (2013).
\bibitem{Loock2016} M. Bergmann and P. van Loock, Quantum error correction against photon loss using multicomponent cat states, Phys. Rev. A \textbf{94}, 042332 (2016).
\bibitem{Heeres2017} R. W. Heeres, Ph. Reinhold, N. Ofek, L. Frunzio, L. Jiang, M. H. Devoret, and R. J. Schoelkopf, Implementing a universal gate set on a logical qubit encoded in an oscillator, Nat. Commun. \textbf{8}, 94 (2017).
\bibitem{Flurin2017} E. Flurin, V. V. Ramasesh, S. Hacohen-Gourgy, L. S. Martin, N. Y. Yao, and I. Siddiqi, Observing Topological Invariants Using Quantum Walks in Superconducting Circuits, Phys. Rev. X \textbf{7}, 031023 (2017).
\bibitem{Loock2006} P. van Loock, T. D. Ladd, K. Sanaka, F. Yamaguchi, K. Nemoto, W. J. Munro, and Y. Yamamoto, Hybrid Quantum Repeater Using Bright Coherent Light, Phys. Rev. Lett. \textbf{96}, 240501 (2006).
\bibitem{Toscano2006} F. Toscano, D. A. R. Dalvit, L. Davidovich, and W. H. Zurek, Sub-Planck phase-space structures and Heisenberg-limited measurements, Phys. Rev. A \textbf{73}, 023803 (2006).


\bibitem{Furusawacomplex} M. Yukawa, K. Miyata, T. Mizuta, H. Yonezawa, P. Marek, R. Filip, and A. Furusawa, Generating superposition of up-to three photons for continuous variable quantum information processing, Opt. Express \textbf{21}, 5529 (2013).
\bibitem{Hofheinz2009} M. Hofheinz \textit{et al.}, Synthesizing arbitrary quantum states in a superconducting resonator, Nature  \textbf{459}, 546 (2009). 
\bibitem{Morin2014} O. Morin, K. Huang, J. Liu, H. Le Jeannic, C. Fabre, and J. Laurat, Remote creation of hybrid entanglement between particle-like and wave-like optical qubits, Nat. Photonics \textbf{8}, 570 (2014).
\bibitem{Jeong2014} H. Jeong \textit{et al.}, Generation of hybrid entanglement of light, Nat. Photonics \textbf{8}, 564 (2014).
\bibitem{Jaromir} J. Fiur\'a\v{s}ek, Efficient representation of purity-preserving Gaussian quantum filters, Phys. Rev. A \textbf{87}, 052301 (2013).
\bibitem{Hage2008} B. Hage, A. Samblowski, J. DiGuglielmo, A. Franzen, J. Fiur\'a\v{s}ek, and R. Schnabel, Preparation of distilled and purified continuous-variable entangled states, Nat. Physics \textbf{4}, 915 (2008).

\bibitem{Andersen} U. L. Andersen, J. S. Neergaard-Nielsen, P. van Loock, and A. Furusawa, Hybrid discrete- and continuous-variable quantum information, Nat. Physics \textbf{11}, 713 (2015).


\begin{thebibliography}{99}

\bibitem{Morin2012} O. Morin, V. D'Auria, C. Fabre, and J. Laurat, High-fidelity single-photon source based on a type II optical parametric oscillator, Opt. Lett. \textbf{37}, 3738 (2012).
\bibitem{LeJeannic2016} H. Le Jeannic \textit{et al.}, High-efficiency superconducting nanowire single-photon detector for quantum state engineering in the near infrared, Opt. Lett. \textbf{41}, 5341 (2016).
\bibitem{Morin2013} O. Morin, C. Fabre, and J. Laurat, Experimentally accessing the temporal mode of traveling quantum light states, Phys. Rev. Lett. \textbf{111}, 213602 (2013).
\bibitem{Huang2015} K. Huang \emph{et al.}, Optical synthesis of large-amplitude squeezed coherent-state superpositions with minimal resources, Phys. Rev. Lett. \textbf{115}, 023602 (2015).


\end{thebibliography}
\end{document}